\documentstyle[12pt,epsfig]{article}
\title{\Large\bf The Neutron Halo in Heavy Nuclei Calculated
with~the~Gogny~Force\thanks{The work is partly supported by the Polish
Committee of Scientific Research under Contract No. 2P03B~011~12}}

\author{      B. Nerlo--Pomorska and K. Pomorski\\
{\it Theoretical Physics Department, Maria Curie-Sk\l odowska University}\\
{\it ul. Radziszewskiego 10, 20-031 Lublin, Poland }\\
\\
\             J. F. Berger and J. Decharg\'{e} \\
  {\it Commissariat \`{a} l'Energie Atomique, Service de Physique
Nucl\'{e}aire }\\
  {\it B.P. 12, 91680 Bruy\`{e}res-le-Ch\^{a}tel, France}}
\date{}

\begin{document}
\maketitle

\begin{abstract}

The proton and neutron density distributions, one- and two-neutron
separation energies and radii of nuclei for which neutron halos are
experimentally observed, are calculated using the self-consistent
Hartree-Fock-Bogoliubov method with the effective interaction
of Gogny.
Halo factors are evaluated assuming hydrogen-like antiproton wave
functions. The factors agree well with experimental data. They are close
to those obtained with Skyrme forces and with the relativistic mean
field approach.

\end{abstract}

\section{Introduction}

In the last years, more and more experimental evidence concerning
neutron distributions in nuclei has become available \cite{Lu98}.
In particular, it has been found that several neutron rich nuclei
display a neutron skin or even a neutron `stratosphere' called the
neutron halo.
Such properties of neutron distributions have been predicted by the
asymptotic density model \cite{Wy93}, the relativistic mean field
theory \cite{Ba95}, and Hartree-Fock calculations with Skyrme forces
\cite{Lu98}.

With the increasing amount of experimental data available, halo factors
can now be used as an additional test of nuclear effective interactions.
Namely, microscopic approaches should be able to reproduce not only
binding energies and charge radii, but also the detail of density
distributions, especially density tails at large distance from the
nucleus center.

The aim of this work is to perform such a test for the effective
interaction proposed by Gogny \cite{De80}.
While many applications have shown the high quality of this force
for describing a wide range of nuclear properties [4--7],
no systematic study in this direction has been made up to now.
The mean-field theory employed with the Gogny force is
applied to nineteen halo nuclei found in \cite{Lu98}. We calculate
neutron and proton density distributions at large distance, one- and
two-neutron separation energies, and the halo factors related to the
probability of annihilation of antiproton from atomic--like orbitals.
These results are compared with experimental data and with Hartree-Fock calculations performed with the SLy4 interaction \cite{Ch97}.

In Section 2, we recall the form of the Gogny effective interaction
and we describe the method by which the results mentioned above are
obtained numerically.
In Section 3, neutron separation energies and density distributions at
large distance from the center of the nucleus are analyzed.
In Section 4, the prescription employed for evaluating halo factors is
given
and the theoretical values obtained are compared with experimental data.
Conclusions and plans for further investigations are presented at the end
of the paper.

\section{Description of the approach}

The Gogny two--body effective nuclear interaction has the
following form \cite{De80} :
\begin{eqnarray}
V_{12} = \sum^2_{i=1} \, \exp\left[-{|\vec r_1-\vec r_2|^2\over
\mu_i^2}\right] \cdot (W_i+B_i\hat P_\sigma - H_i\hat P_\tau - M_i\hat
P_\sigma\hat P_\tau)+ \nonumber\\
+ t_3 (1 + x_0\hat P_\sigma)\,\delta (\vec r_1 - \vec r_2)
\left[\rho\left({\vec r_1+\vec r_2\over 2}\right)\right]^\gamma +\\
 + iW_{\rm LS}(\vec\sigma_1 + \vec\sigma_2) \cdot
\stackrel{\leftarrow}{\nabla}_{12}
    \times \delta(\vec r_1-\vec r_2) \vec\nabla_{12} + V_{\rm Coul.}\,,
\nonumber
\end{eqnarray}
The first line represents two finite range terms ($i=1,2$), with the
usual superposition of Wigner, Bartlett, Heisenberg and Majorana
spin-isospin
contributions.
$\vec r_i$ is the space coordinate of nucleon $i$, and  $\hat P_\sigma$
and $\hat P_\tau$ are the exchange operators of  spin and isospin
variables, respectively.
The second line of eq. (1) describes a two-body zero-range
density--dependent interaction.
The last line contains a two-body zero-range spin--orbit term and the
Coulomb potential between protons.
Here,  $\vec\nabla_{12}=\vec\nabla_1-\vec\nabla_2$, and $\vec\sigma_i$
is twice the spin operator of nucleon $i$.

The set of parameters adopted since 1983, called D1S \cite{Be84}, is:\\
\begin{center}
\begin{tabular}{ll}
$\mu_1$ = 0.7 fm            &   $\mu_2$ = 1.2 fm      \\
$W_1$ = -1720.3 MeV         &   $W_2$ = 103.639 MeV   \\
$B_1$ = 1300 MeV            &   $B_2$ = -163.483 MeV  \\
$H_1$ = -1813.53 MeV~~~~~~~~~~~~        &   $H_2$ = 162.812 MeV   \\
$M_1$ = 1397.60 MeV         &   $M_2$ = -223.934 MeV  \\
$t_3$ = 1390.60 MeV fm$^{3(1+\gamma)}$        &   $x_0$ = 1         \\
$\gamma$ = 1/3              &   $W_{LS}$ = 130 MeV fm$^5$ \\
\end{tabular}
\end{center}

Hartree-Fock-Bogoliubov (HFB) self-consistent calculations have been
carried out with this interaction, with the purpose of extracting binding
energies and nucleon density distributions.
Similar results employing the Hartree-Fock (HF) method, i.e. ignoring
pairing correlations, have also been derived in order to compare them
with those obtained with the Skyrme HF method.
All calculations have been performed in spherical symmetry. The
method used is described in detail in Ref. \cite{De80}.
The assumption of a spherical mean-field may seem inappropriate for
those halo nuclei which ground states are deformed.
However, since halo factors depend essentially on the slope of the 
density at
very large distance from the nuclear surface, such a simplification
should be sufficient for a first estimate of their magnitudes.
In the same spirit, the effect on nuclear densities of correlations
beyond the mean-field approximation, as those coming from oscillations
around a fixed shape or from large amplitude collective vibrations
\cite{Go79,Gi76} has been ignored in the present study.
One should nevertheless keep in mind that neglecting the increase of
neutron and proton radii caused by deformation and correlations
should lead to a slight underestimation of halo factors in most nuclei.

Our calculations are performed by expanding the HFB one quasi-particle
states on finite harmonic oscillator (HO) bases.
A crucial point in this method is to carefully choose the two parameters:
number of shells (N$_{\rm MAX}$) and harmonic oscillator frequency $\hbar\omega$,
on which these bases depend.
This is especially needed in the present study where the behaviour of
single-particle wave-functions at large distance from the center of the
nucleus has to be accurately determined. The method we have employed
consists of choosing, for each value of N$_{\rm MAX}$, the $\hbar\omega$  value
that minimizes the HFB binding energy. Then N$_{\rm MAX}$ is increased until
convergence of the HFB binding energy is obtained.

Application of this method to $^{58}$Ni is illustrated in Figs. 1, 2 and 3.
Fig.1 displays the variations of the HFB energy ($B$) with $\hbar\omega$ for
N$_{\rm MAX}$=16.
The energies of the minima of the curves $B(\hbar\omega)$ are plotted
in Fig.2 as functions of N$_{\rm MAX}$. One observes that convergence of the HFB
binding energy is obtained for N$_{\rm MAX}$=16 in this nucleus.
As shown in Fig.3, a well defined energy minimum is then obtained in
the $(\hbar\omega, N)$ plane.
This procedure has
been employed for the nineteen halo--nuclei considered in the present
study.

In order to illustrate the influence of the HO basis parameters on
densities, we have plotted in Fig. 4 the logarithm of the neutron
density distribution $\log\rho_n$ of $^{58}$Ni versus the radial distance
$r$ for a few values of $\hbar\omega$.
One can see that $\rho_n$ is sensitive to the chosen $\hbar\omega$ only in
the peripheral region $r >$ 8 fm, a region where $\rho_n$ is smaller than
$10^{-10}$ fm$^{-3}$.

\section{Results}

The parameters of the HO bases employed for all the halo--nuclei studied in
the present work are gathered in Table~1.
The HO frequency $\hbar\omega$
corresponding to the minimal HFB energy and the number of shells N$_{\rm MAX}$ 
above which the energy does not change any more are shown for each nucleus.
The HFB energy and the one-- and two--neutron separation energies $S_n$
and $S_{2n}$ are also listed.
The separation energies have been found by subtraction of the HFB
energies of neighbouring isotopes
\begin{equation}
 S_n(Z,N) = B(Z,N) - B(Z,N-1)\,\,,
\end{equation}
\begin{equation}
 S_{2n}(Z,N) = B(Z,N) - B(Z,N-2) \,\,,
\end{equation}
Odd isotope energies have been calculated using the blocking version of the HFB
theory, as described in Ref. \cite{De80}. In each nucleus, the blocked
quasi-particle state has been chosen as the one having the experimentally
known spin nearest to the neutron Fermi surface.
Quadrupole deformations $\beta_2$ taken from Ref. \cite{Pe92} are 
given in Table~1 in order to indicate which nuclei are deformed in their 
ground states. Let us recall that
all nuclei are considered as spherical in our calculation.
The last two columns of Table~1 display calculated and experimental halo
factors, the meaning of which will be explained in Section 4.

  The one-- and two--neutron separation energies of these nuclei are compared
with the experimental values \cite{Au95} in Figs. 5 and 6. In these figures, the
differences between experimental and theoretical separation energies are
plotted as functions of the proton number $Z$. 

Calculated neutron separation energies agree with experiment within
$\pm 1$ MeV in spherical nuclei, except for $^{48}$Ca, $^{96}$Ru,
$^{130}$Te and $^{144}$Sm. In these nuclei, the neutron Fermi level is
located in the vicinity of a major shell gap, where the single-particle
level density is underestimated in the present mean-field approach.
A proper description of separation energies in nuclei near closed
shells would require to take into account correlations beyond the
mean-field -- as RPA ground state correlations -- and the induced
spectroscopic factors and single-particle level displacements.
Let us also note that, in the calculation of $S_n$s, odd nuclei have
been assumed to be spherical, which leads to a systematic
overestimation of theoretical values. 
In the case of deformed nuclei, one observes that $S_{n}$s usually are
overestimated, while $S_{2n}$s often are underestimated with the
present spherical approach.

Calculations of proton and neutron density distributions have been
performed for the nineteen halo nuclei using the HF and HFB procedures
with the Gogny interaction.
For the purpose of comparison, similar results have been derived also
with the SLy4 parametrization \cite{Ch97} of the Skyrme force \cite{Sk59} 
using the HF method.

The proton and neutron density distributions $\rho_{n(p)}$ for $^{58}$Ni
are presented in Fig. 7. The Skyrme interaction gives proton and neutron
densities lower by $\simeq$ 10\% in the nucleus interior and, consequently
slightly larger proton and neutron radii than the Gogny force.
However, the tails of the densities obtained within the
two approaches do not differ significantly from each other. Let us recall
here that the parameters of the Gogny force have been determined in order
to allow for the inclusion of ground state correlations in the description
of one-body observables. As a consequence, nuclear radii
are expected to be accurately reproduced by the Gogny force only when
correlations beyond the mean-field are included.

The difference between the proton and the neutron distributions at
large distance is important for the investigation of halo effects.
In order to illustrate the detailed structure of these distributions,
several functions of the densities $\rho_{n(p)}$ which enter halo factors
(see Section 4) have been plotted.
A complete set of such results is displayed in Fig. 8 for $^{160}$Gd as
an example.  Similar plots for the remaining eighteen nuclei are shown
in the next eight figures.

The upper part of Fig. 8 represents the density distributions for
protons (solid lines) and neutrons (dashed lines) obtained with the Gogny
force without pairing correlations (GoHF), with pairing correlations
included
(GoHFB) and with the Skyrme force SLy4 (SkHF) without pairing
correlations.
The densities are plotted up
to $r$ = 12~fm, a radial distance above which they are lower than
$10^{-10}$ fm$^{-3}$. The middle row shows the logarithms of these 
densities, and the leftmost diagram of
the lowest row the difference $\log \rho_{n} - \log \rho_{p}$.

One can see that the densities computed with and without pairing
correlations are very close to each other, although the contribution
of pairing to the binding energy is of the order of 10 MeV.

More significant differences can be noticed between the results
obtained with the SLy4 and the Gogny forces (both without pairing),
especially for large $r$ and near $r=0$.

It is interesting to analyze how the different single-particle states
contribute to the total density $\rho$. Denoting by $\rho_\nu$ the
contribution of single-particle state $\nu$ :
\begin{equation}
 \rho = \sum_\nu \rho_\nu
\end{equation}
the relative contributions $\rho_\nu / \rho$ of the occupied
single-particle orbits are plotted in the rightmost diagram of the
lowest row in Fig. 8. These results have been obtained from HF calculations
with the Gogny force.
One observes that all single particle states except one,
contribute more or less the same amount in the whole region $r= 0-12$ fm.
At large distance and for $r \simeq 0$,
the  $f_{7/2}$ one-neutron state strongly dominates
all the other ones. This clearly indicates that the halo is a single
particle  effect in this nucleus, which confirms earlier results
obtained with Skyrme forces \cite{Lu98} and with the relativistic mean
field theory \cite{Ba95}.

The middle plot of the lowest row in Fig. 8 shows the function
$(\rho_n-\rho_p)\cdot r^2/(N-Z)$ calculated within the three approaches
GoHFB (solid line), GoHF (dashed line), and SkHF (dotted line). This
function
is strongly correlated with the magnitude of halo factors.

Similar results for the other eighteen halo nuclei listed in Table 1
are shown in the next eight figures: 9a, 9b, 10a, 10b, 11a, 11b, 12a,
and 12b. Figures ''a'' correspond to the lighter Ca-Sn nuclei
and figures ''b''  to the heavier Sn-U. Every multiplot presents the
same quantity versus $r$ for nine different nuclei.

In Fig. 9, the differences $\log \rho_n - \log \rho_p$ between the
logarithms of the neutron and proton densities are shown.
The solid lines have been obtained with the Gogny force
(GoHF), and the dashed ones with the SLy4 (SkHF) interaction.
The difference between the proton and neutron densities grows with $r$
and is larger for the Skyrme force in lighter nuclei (9a) while, for the
heavier ones (9b), the Gogny force gives similar or larger density
differences than SLy4.

Fig. 10 shows the contributions $\rho_\nu$ of neutron (solid lines) and
proton (dashed lines) occupied single particle orbitals $\nu$ to
the whole density $\rho$.
One can see that in most cases only one neutron state determines the
magnitude
of the density tail.
For lighter nuclei (10a), only $^{116}$Cd and $^{100}$Mo have more than
one neutron state contributing to the tail of the density.
In heavier nuclei, Fig. 10b, the Sn and Te isotopes show a similar
behaviour, with two neutron orbitals contributing to the density tail.
A very interesting situation is found in $^{144}$Sm where the
contribution that dominates corresponds to a proton state.
Therefore a proton halo can be expected in this
nucleus. This is in line with experimental data \cite{Lu98}. 
Unfortunately this is
not the case for $^{106}$Cd, which also has a proton-rich nuclear
stratosphere.
For every plot in Fig. 10, the quantum numbers $lj$ of the orbital for
which
the contribution $\rho_\nu$ is maximum at large distance is indicated.
The single particle character of the nuclear periphery has also been found
in HF calculations with the Skyrme force SkIII \cite{Lu98} and
within the relativistic mean field theory \cite{Ba95}.

The two Fig. 11a,b show the function $(\rho_n-\rho_p)r^2/(N-Z)$ versus
$r$, for the eighteen halo nuclei.
This quantity directly enters the integral for the halo factor, Eq. (6),
and determines the neutron or proton halo in the nuclear periphery. HF
results are shown both for the Gogny (solid line) and the SLy4 (dashed
line) forces. It is apparent on these curves that the Skyrme force
yields a slightly larger neutron halo than the Gogny interaction.

Figs. 12a and 12b illustrate the nuclear skin effect. The $r^2$--weighted
difference between the average neutron ($\rho_n/N$) and proton
($\rho_p/Z$) single particle densities calculated with the Gogny force
is plotted for each nucleus as a functions of $r$. One can see that
this quantity strongly oscillates in the surface region, with
variations that significantly differ in amplitude and shape, depending on
the nucleus considered.

\section{Halo factor}

In experiments probing the nuclear periphery using the formation of
antiproton atoms,  antiprotons
are catched on hydrogen--like atomic orbitals and then annihilated on
the outer tail of the proton or neutron density.
The halo factor is defined as \cite{Lu98} :
\begin{equation}
 f \simeq {Z\sum\limits_s\Gamma^s_n\over N\sum\limits_s\Gamma^s_p}\,,
\end{equation}
where the summation goes over  all the antiprotonic states with
meaningful annihilation widths $\Gamma^s_{n(p)}$ on proton ($p$) or
neutron ($n$) and $s$ denotes the atomic state of the antiproton.

The widths are calculated by integrating the density of one kind of
nucleon with the square of the antiproton wave-function $\psi^s(r)$ :
\begin{equation}
\Gamma^s_{n(p)} = \int\rho_{n(p)}|\psi^s_{n(p)}(r)|^2 P(r)\,r^2 dr\,.
\end{equation}
$\psi^s(r)$ is taken as the solution of the Schr\"{o}dinger equation for a
hydrogen-like antiprotonic atom.
The nucleus is assumed to be spherical. Let us point out that the nuclear
radius is much smaller than typical antiproton orbital radii.

The factor $P(r)$ describes the probability that the annihilation products
of the antiproton will not be absorbed by the nucleus. This factor, which
should depend on the atomic state of the antiproton, will be assumed here
to be independent of it.
$P(r)$ takes into account the deep hole creation probability
$P_{\rm dh}(r)$ and
the pion escaping probability $P_{\pi,{\rm esc}}(r)$. For the ground state
nuclear periphery, only cold annihilation is important. Hot annihilation
has to be eliminated from the widths $\Gamma$, which leads to the $P(r) =
{\rm constant}$ approximation.

In Fig. 13 the halo factors obtained with the Gogny and Skyrme forces are
compared with the experimental data taken from Ref. \cite{Lu98}.
The variations of experimental halo factors from nucleus to nucleus are
satisfactorily reproduced by the two theoretical calculations. In most cases,
the Gogny force gives smaller halo factors than the Skyrme interaction,
which is consistent with the fact that the Gogny D1S parameterization
underestimates radii when correlations beyond the mean field are neglected.
This is especially true in the five heavier nuclei which all are deformed.
The halo factors derived with SLy4 appear in good quantitative agreement
with experimental data. In particular, the unusually large halo factor in
$^{176}$Yb is well reproduced. 
Let us note that both effective interactions significantly
underestimate the halo factors of the two Te isotopes,
which indicates that the halo structure of these two spherical nuclei
is not properly described by the present microscopic calculations.
One reason for these discrepancies may be that correlations beyond the
mean-field approximation are necessary for a correct description of
neutron densities at large distance in these nuclei.

\section{Conclusions}

Hartree--Fock--Bogoliubov calculation in spherical symmetry have been
performed with the Gogny force for nineteen halo nuclei.  The
parameters of the HO bases employed -- maximum number of shells N$_{\rm
MAX}$ and frequency $\hbar\omega$ -- have been carefully chosen in
order to ensure the convergence of the total binding energy, and to
describe the large distance behaviour of nucleon density
distributions.

The following conclusions can be drawn from the present investigation:

\begin{enumerate}

\item One-- and two-- neutron separation energies are reproduced with the Gogny force within $\pm$ 1 MeV in most of the spherical nuclei studied. 
In deformed nuclei, the spherical symmetry assumption leads to an
underestimation of $S_{2n}$s and an overestimation of $S_{n}$s that can
reach 3 MeV.

\item The Gogny force D1S gives an overall satisfactory account of the
nuclear periphery which could probably be improved by taking into
account deformation effects and, to a lesser extent, correlations
beyond the mean field.  Inclusion of pairing correlations in the
self-consistent calculation induces only weak changes in the proton and
neutron density distributions, and has a negligible effect on halo
factors.

\item The variations of halo factors from nucleus to nucleus obtained
with the Gogny force and the SLy4 interaction are in good agreement
with experimental data.  From the present spherical mean-field
calculations, SLy4 gives a larger neutron halo effect than the
Gogny force, in better agreement with experiments.

\end{enumerate}

As almost half of experimentally known halo--nuclei are well deformed,
a test of the influence of including nuclear deformation
in the self-consistent calculations is clearly needed.
In the case of the Gogny force, a similar test concerning the effect of
the ground state correlations associated with oscillations of the
mean-field -- large amplitude collective motion in soft nuclei, RPA
ground state correlations in rigid nuclei --  and of possible shape
coexistence phenomena should also be performed. This is left for future work.

\newpage
{\Large{\bf Table captions:}}
\vspace{5mm}
\begin{itemize}
\item[1.] The results of the Hartree-Fock-Bogoliubov calculation with
the Gogny D1S force for 19 halo nuclei \cite{Lu98}. The harmonic oscillator 
basis parameters: the shell number N$_{MAX}$ and the oscillator frequency 
$\hbar\omega$, the binding energies $B$, one $S_n$ and two $S_{2n}$
neutron separation energies, the equilibrium deformation $\beta_2$
and the theoretical $f$ and experimental halo factors are listed in the table.
\end{itemize}

{\Large{\bf Figure captions:}}
\vspace{5mm}
\begin{itemize}

\item[1.] Hartree-Fock energy (crosses) obtained with the Gogny force
in $^{58}$Ni
versus the frequency ($\hbar\omega$) of the harmonic oscillator
basis. The theoretical points are interpolated by 2$^{nd}$ order polynomial
in $\hbar\omega$.

\item[2.] Hartree-Fock energy (crosses) obtained with the Gogny force
in $^{58}$Ni as a function of the number (N$_{\rm MAX}$)
of harmonic oscillator shells included in the basis.

\item[3.] Contours of the Hartree-Fock energy (crosses) obtained with the
Gogny force in $^{58}$Ni in the (N$_{\rm MAX}$, $\hbar\omega$) plane. 
N$_{\rm MAX}$ is
the number of shells included in the harmonic oscillator basis
and $\hbar\omega$ is the harmonic oscillator frequency.

\item[4.] Logarithms of the neutron density distribution $\rho_n$ as
functions of the radial distance $r$ in  $^{58}$Ni. The different curves
correspond to different values of
the harmonic oscillator basis frequency $\hbar\omega$.

\item[5.] Difference between theoretical (Gogny) and experimental \cite{Au95}
one--neutron separation energies $S_n$ for experimentally known
halo--nuclei.

\item[6.] Same as Fig.5 for two--neutron separation energies $S_{2n}$.

\item[7.] Proton (dashed and dotted lines) and neutron (solid, thin dashed
     lines) density
distributions obtained from Hartree-Fock calculations with the Gogny force
(GoHF) and with the Skyrme SLy4 force (SkHF), versus the radial distance $r$.

\item[8.] Density distribution results obtained in $^{160}$Gd.
First row:
Density distributions for protons $\rho_p$ (solid lines) and neutrons $\rho_n$
(dashes lines) obtained from Hartree-Fock calculations with the Gogny force D1S
(first column), from Hartree-Fock-Bogoliubov calculations with the Gogny force
(second column), and from Hartree-Fock  calculations with the Skyrme force
SLy4 (third column).
Second row: Logarithms $\log \rho_{n(p)}$ of the densities shown in the
first row.
Third row, first column: the difference $\log\rho_n - \log\rho_p$
between the logarithms of the neutron
and proton densities; second column:
the function $(\rho_n - \rho_p)r^2/(N-Z)$ obtained by applying
the Hartree-Fock method with the Gogny force D1S (GoHF, dashed line),
the Hartree-Fock-Bogoliubov method with the Gogny force (GoHFB, solid line)
and the Hartree-Fock method with the Skyrme force SLy4
(SkHF, dotted line);
third column: single-particle contributions $\rho_\nu$ to the
density $\rho$ for protons (solid lines) and
neutrons (dashed lines).
The $lj$ quantum numbers of the orbit with largest $\rho_\nu$ at $r =
12$~fm are indicated in each plot.

\item[9a.] The differences $\log\rho_n - \log \rho_p$ obtained
from Hartree-Fock calculations with the Gogny force D1S (GoHF -- solid lines),
and with the Skyrme force SLy4 (SkHF -- dashed lines)
for the nine lighter halo--nuclei $^{48}$Ca to $^{112}$Sn.

\item[9b.] Same as Fig. 9a
for the nine  heavier halo--nuclei $^{124}$Sn to $^{238}$U.

\item[10a.] Relative contributions $\rho_\nu / \rho$ of the single particle
proton (dashed lines) and neutron (solid lines) states $\nu$ to the
total density $\rho$ as functions of the radial distance $r$
for the nine lighter halo--nuclei $^{48}$Ca to $^{112}$Sn.

\item[10b.] Same as Fig. 10a for
the nine  heavier halo--nuclei $^{124}$Sn to $^{238}$U.

\item[11a.] The function ($\rho_n - \rho_p)r^2/(N-Z)$ obtained
from Hartree-Fock calculations with the Gogny force D1S (solid lines),
and with the Skyrme force SLy4 (dashed lines)
versus the radial distance $r$
for the nine lighter halo--nuclei $^{48}$Ca to $^{112}$Sn.

\item[11b.] Same as Fig. 11a
for the nine  heavier halo--nuclei $^{124}$Sn to $^{238}$U.

\item[12a.] The function $\left({\rho_n/N} - {\rho_p/Z}\right)r^2$
obtained from Hartree-Fock-Bogoliubov calculations with the Gogny force
for the nine lighter halo--nuclei $^{48}$Ca to $^{112}$Sn.

\item[12b.] Same as Fig 12a
for the nine  heavier halo--nuclei $^{124}$Sn to $^{238}$U.

\item[13.] The halo factors of experimentally known halo--nuclei obtained
from Hartree-Fock-Bogoliubov calculations with the Gogny force D1S (stars 
joined by the thin dashed lines) and from Hartree-Fock calculations 
with the Skyrme forces SLy4 (crosses joined by dashed lines),
compared with experimental data (plusses wit errorbars joined by solid
lines) from Ref. [1].
\end{itemize}

\newpage
\begin{center}
{\bf\Large Table 1}
\vspace{1cm}

\begin{tabular}{|l|c|c|c|c|c|c|c|c|}
\hline
Nucleus & N$_{\rm MAX}$ & $\hbar\omega$ & $|B|$ & $S_n$ & $S_{2n}$ &
$\beta_2$ \cite{Pe92} &  $f$ & $f_{\rm exp} \cite{Lu98}$ \\
\hline
 -- & -- & MeV & MeV & MeV & MeV & -- & -- & -- \\
\hline
$^{48}$Ca & 14 & 16.442 & 416.838 & 11.574 & 15.735 & 0 & 1.77 &
2.35$\pm$0.35 \\
$^{58}$Ni & 16 & 15.470 & 505.161 & 11.663 & 21.562 & 0 &
1.10 & 1.30$\pm$0.2 \\
$^{96}$Zr & 16 & 13.479 & 825.345 & ~7.958 & 12.202 & --0.19  & 2.04 &
3.7$\pm$0.6\\
$^{96}$Ru & 16 & 15.070 & 827.972 & ~9.616 & 17.366 & 0 & 1.14 &
1.10$\pm$0.2 \\
$^{100}$Mo & 18 & 12.962 & 856.871 & ~8.070 & 12.980 & --0.27  & 2.53 &
4.10 \\
$^{104}$Ru & 18 & 14.256 & 888.612 & ~9.233 & 14.001 & --0.28
& 2.24  & 4.30\\
$^{106}$Cd & 16 & 14.254 & 903.668 & 10.688 & 18.138 & 0 & 1.38&
0.60$\pm$0.1\\
$^{112}$Sn & 16 & 14.212 & 952.850 & 10.962 & 18.438 & 0 & 1.56 & \\
$^{116}$Cd & 16 & 13.761 & 984.182 & ~8.861 & 14.793 & --0.28
& 2.29 &  \\
$^{124}$Sn & 16 & 13.261 & 1049.635 & ~9.113 & 14.683 & 0&2.47 & \\
$^{128}$Te & 16 & 13.558 & 1078.166 & ~9.643 & 15.568 & 0 & 2.02 &
4.3$\pm$1.1\\
$^{130}$Te & 16 & 13.205 & 1094.724 & 11.008 & 16.558 & 0 & 2.32 &
4.2$\pm$0.4\\
$^{144}$Sm & 18 & 14.141 & 1197.497 & 12.838 & 20.981 & 0 & 
2.55 & $<$0.5\\
$^{148}$Nd & 16 & 11.898 & 1216.816 & ~8.458 & ~9.831 & 0.23
& 2.40 & 4.8$\pm$0.9\\
$^{154}$Sm & 18 & 12.351 & 1253.112 & ~8.674 & 10.659 & 0.28 &
2.56 & 2.2$\pm$0.4 \\
$^{160}$Gd & 18 & 12.222 & 1290.583 & ~7.406 & 10.816 & 0.31 & 3.27 &
5.8$\pm$1.9\\
$^{176}$Yb & 18 & 10.182 & 1397.546 & ~8.999 & 12.423 & 0.31
& 3.77 &
8.0$\pm$0.6 \\
$^{232}$Th & 18 & ~9.947 & 1747.601 & ~7.724 &  9.540 &
0.21 & 4.00 & 5.4$\pm$0.8 \\
$^{238}$U & 18 & 10.227 & 1778.929 & ~7.893 & 10.059 & 0.24
& 3.70 &
5.8$\pm$0.8\\
\hline
\end{tabular}
\end{center}

\newpage
\pagestyle{empty}

\framebox{Fig.~1}

\vspace{0.5cm}
\epsfig{file=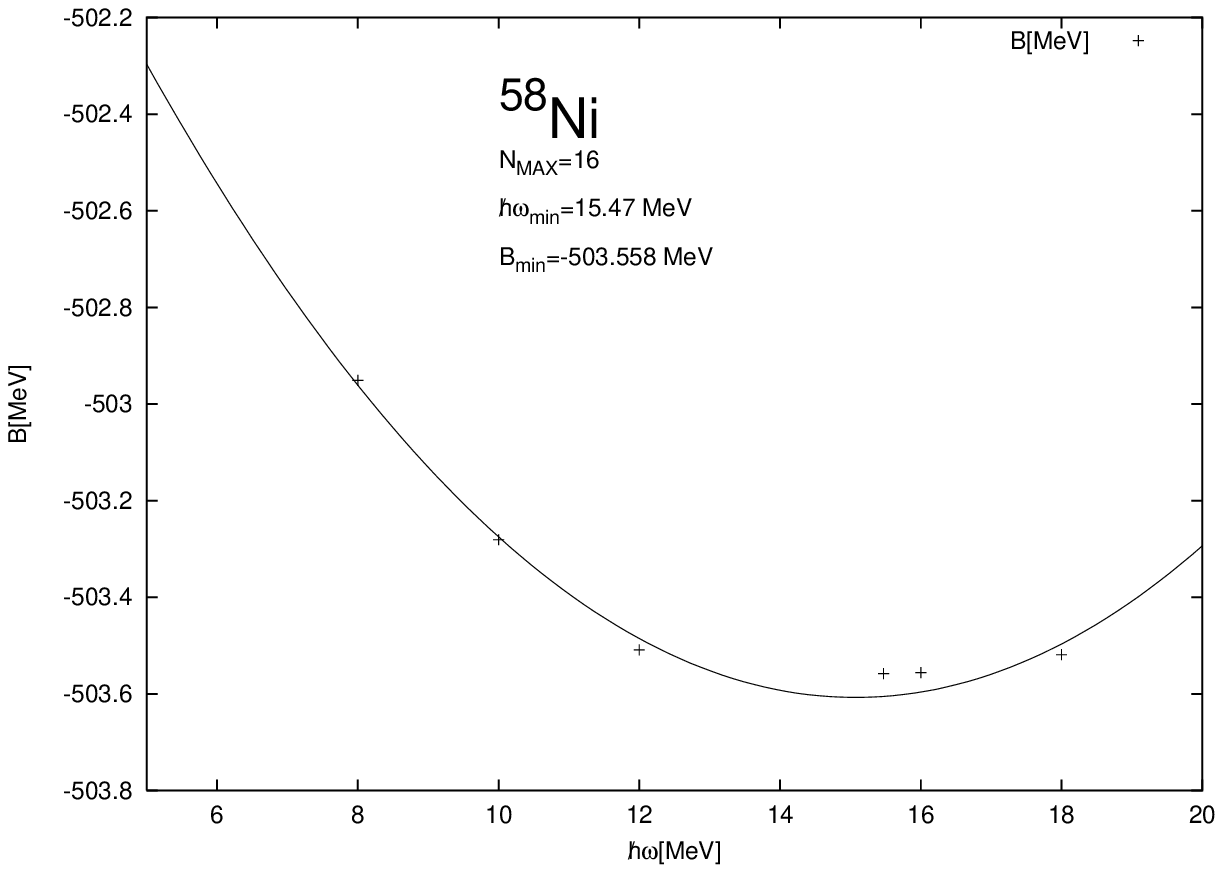,width=140mm}
\newpage

\framebox{Fig.~2}

\vspace{0.5cm}
\epsfig{file=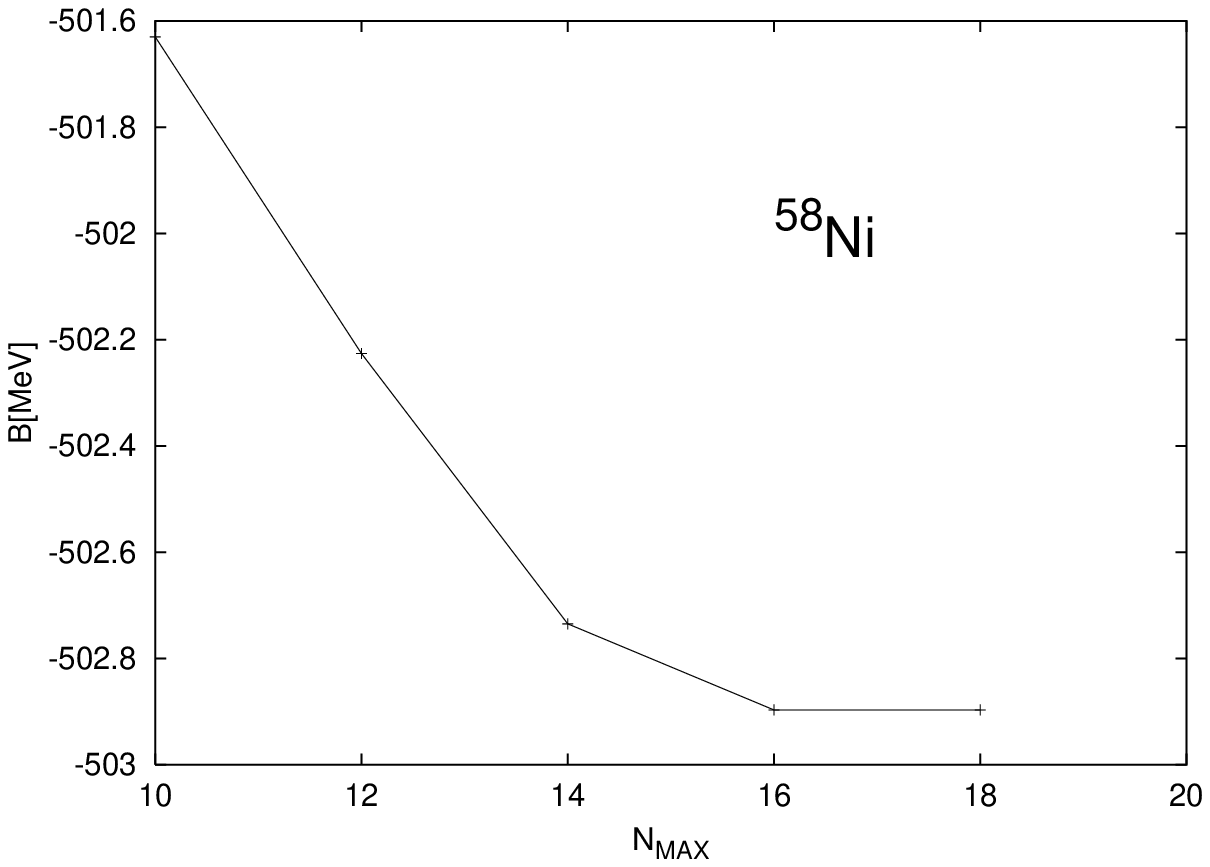,width=140mm}
\newpage

\framebox{Fig.~3}

\vspace{0.5cm}
\epsfig{file=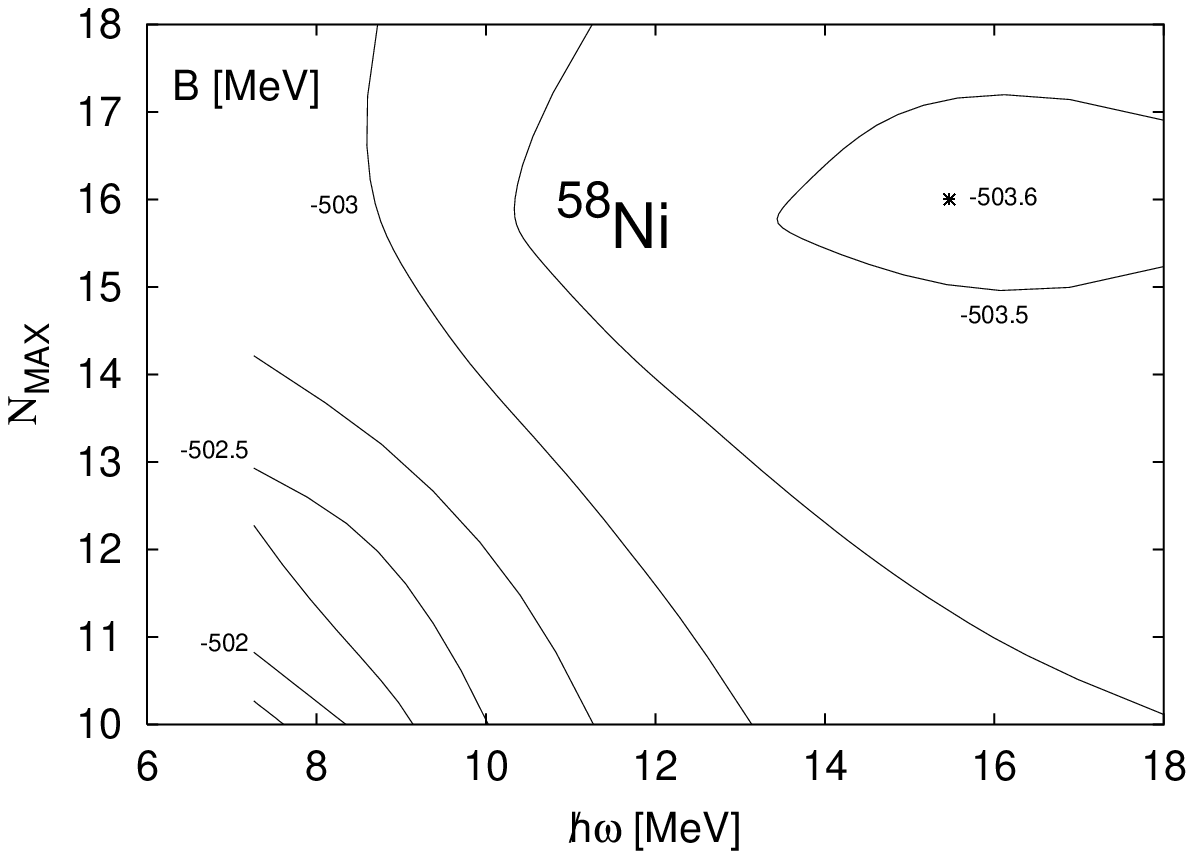,width=140mm}

\newpage

\framebox{Fig.~4}

\vspace{0.5cm}
\epsfig{file=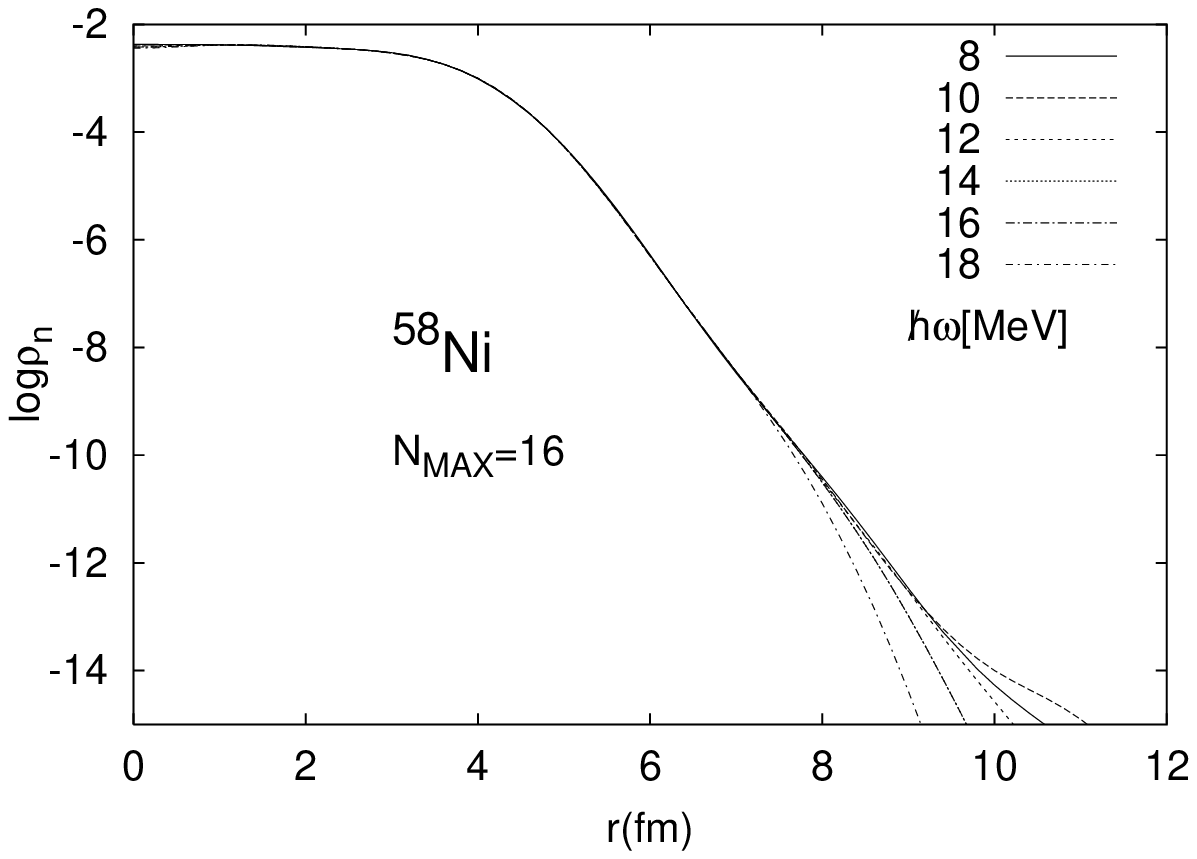,width=140mm}

\newpage

\framebox{Fig.~5}

\vspace{0.5cm}
\epsfig{file=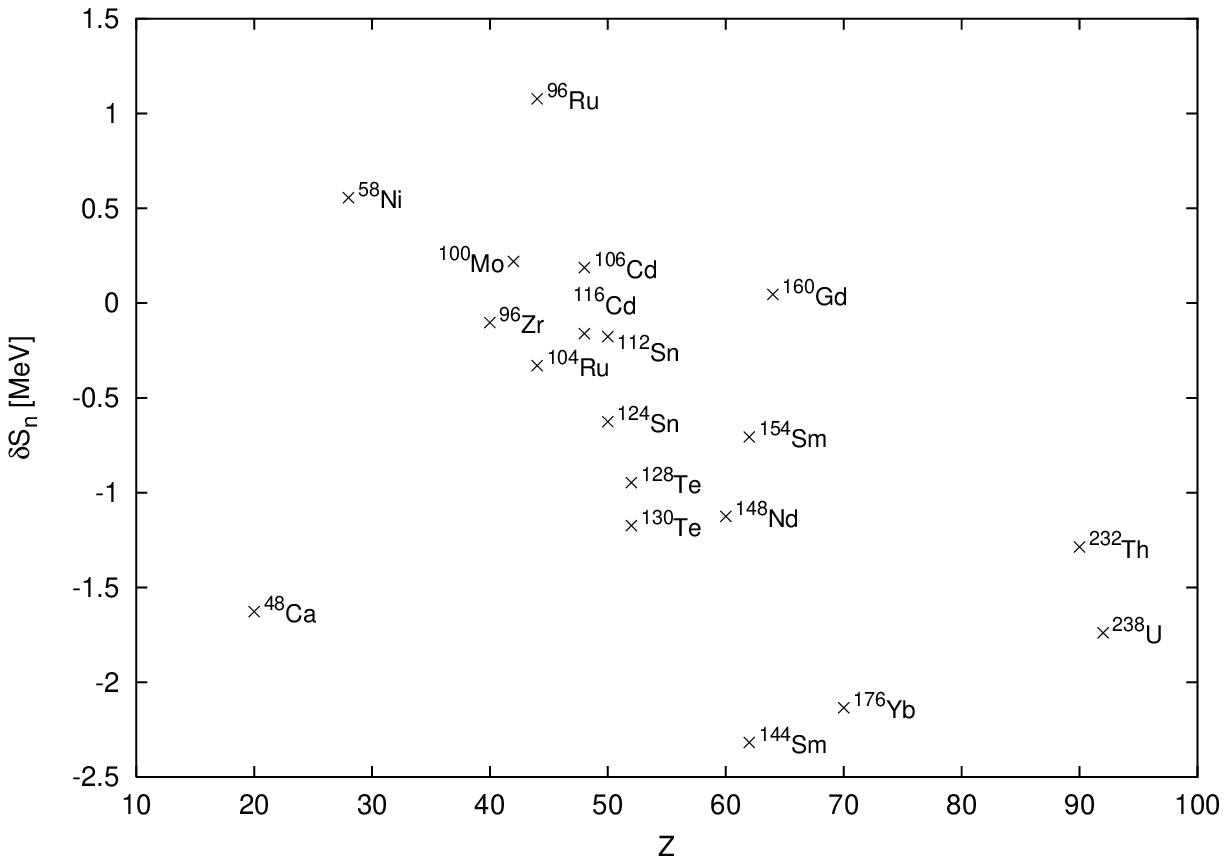,width=140mm}
\newpage

\framebox{Fig.~6}

\vspace{0.5cm}
\epsfig{file=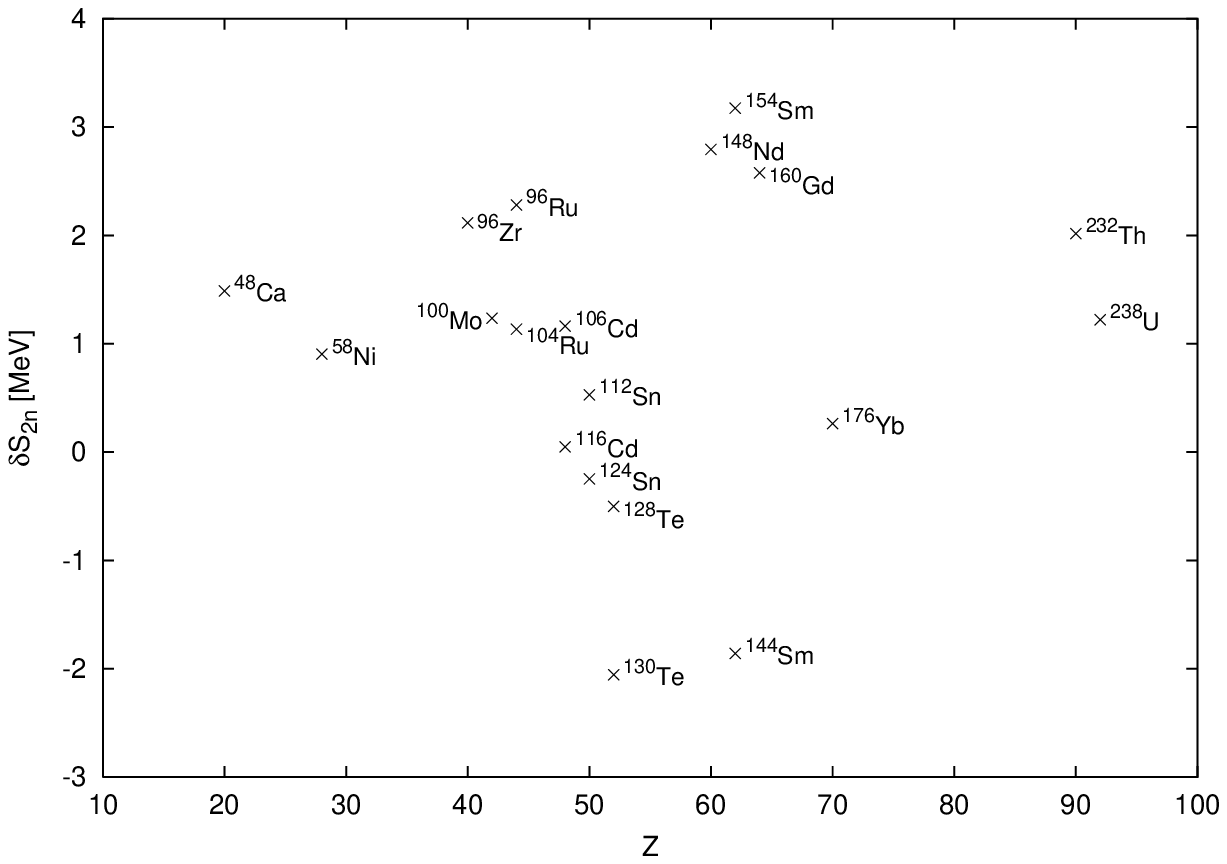,width=140mm}

\newpage

\framebox{Fig.~7}

\vspace{0.5cm}
\epsfig{file=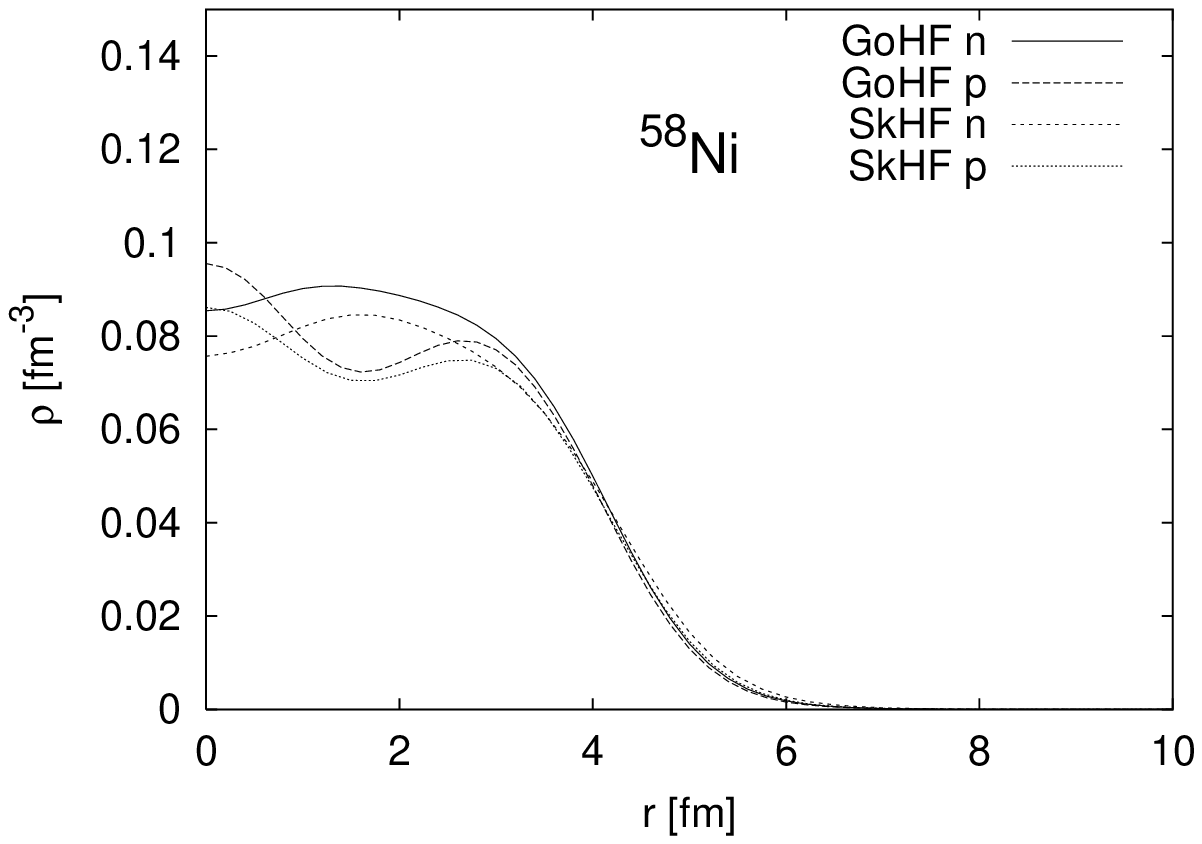,width=140mm}

\newpage

\framebox{Fig.~8}

\vspace{0.5cm}
\epsfig{file=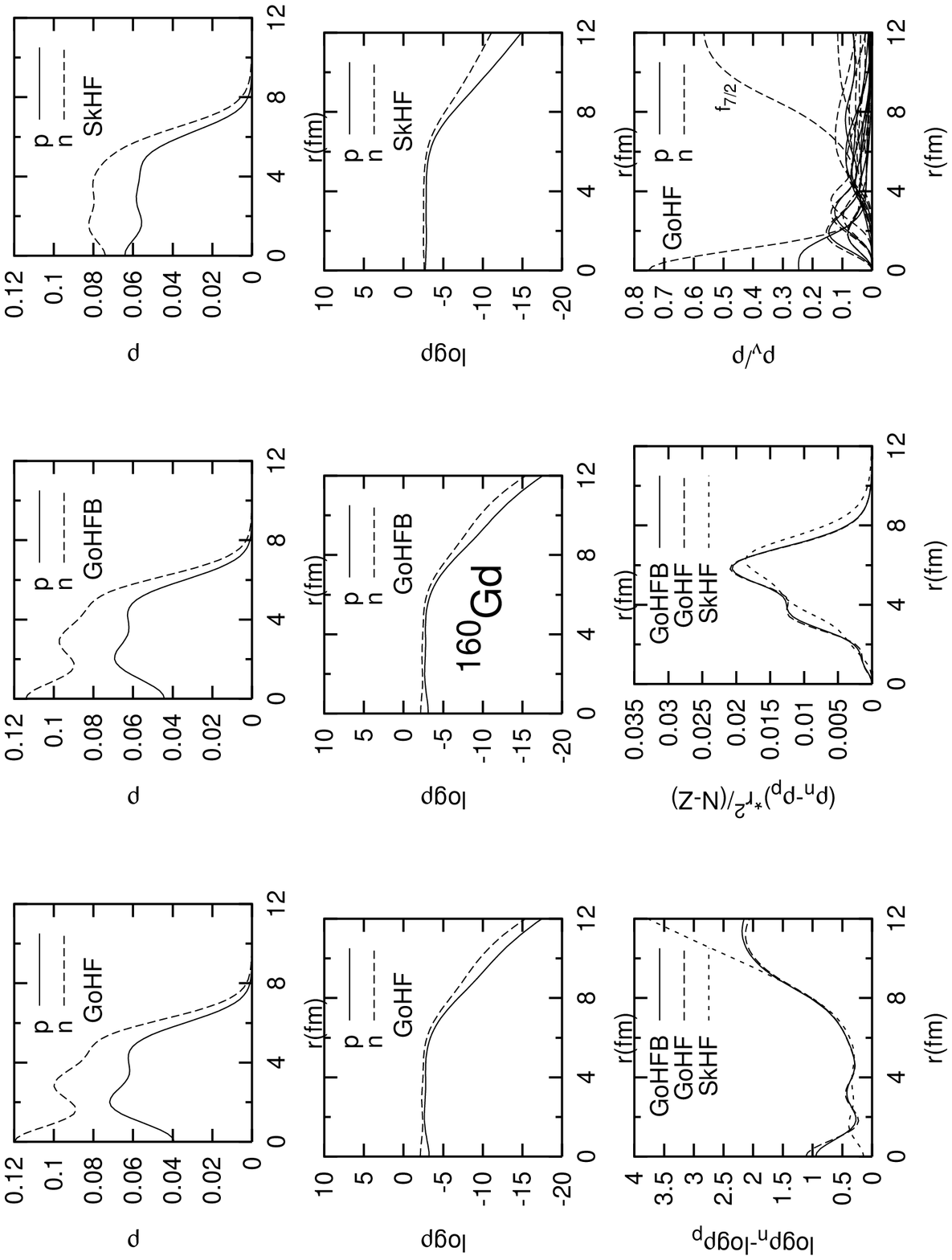,width=140mm}
\newpage

\framebox{Fig.~9a}

\vspace{0.5cm}
\epsfig{file=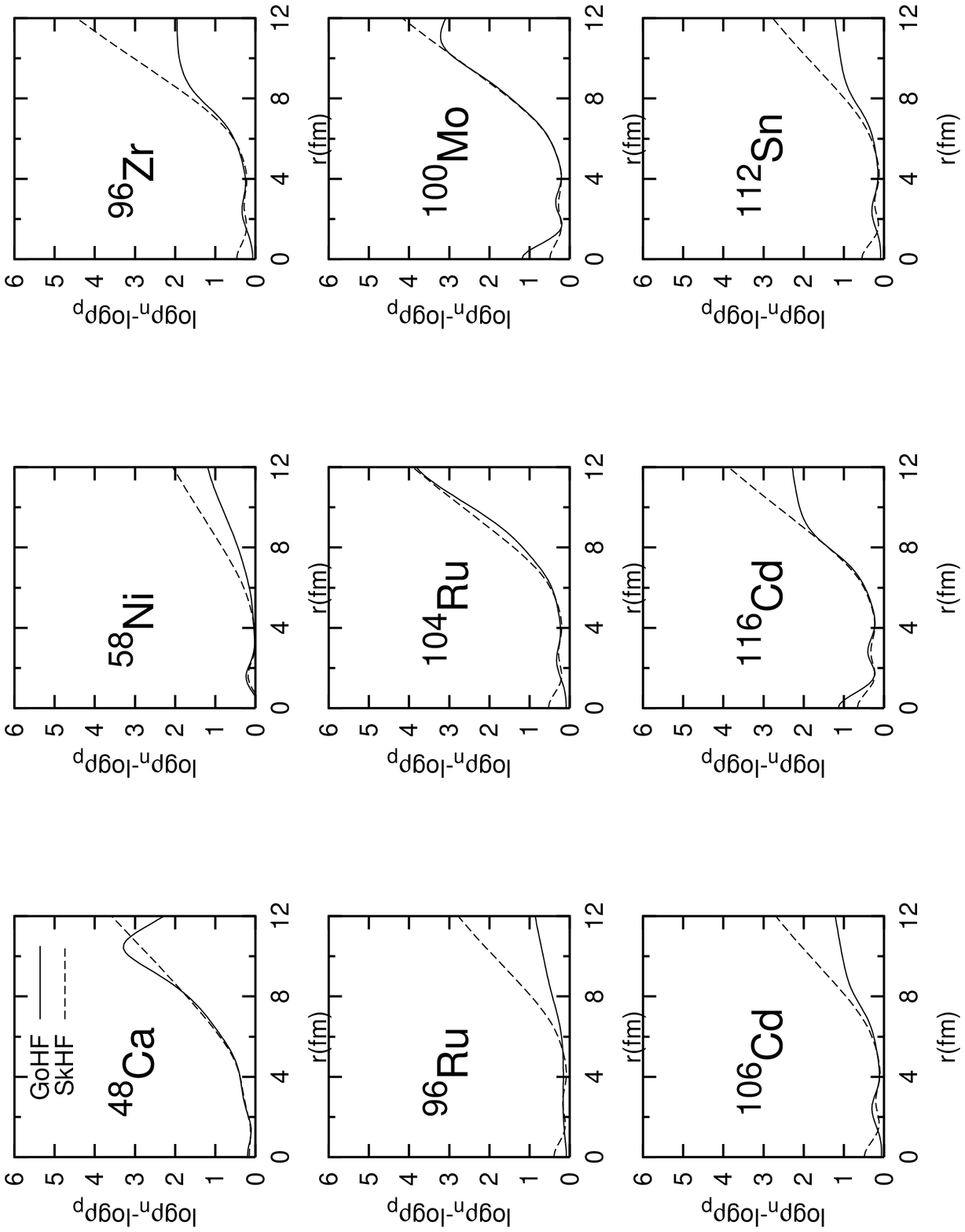,width=140mm}

\newpage

\framebox{Fig.~9b}

\vspace{0.5cm}
\epsfig{file=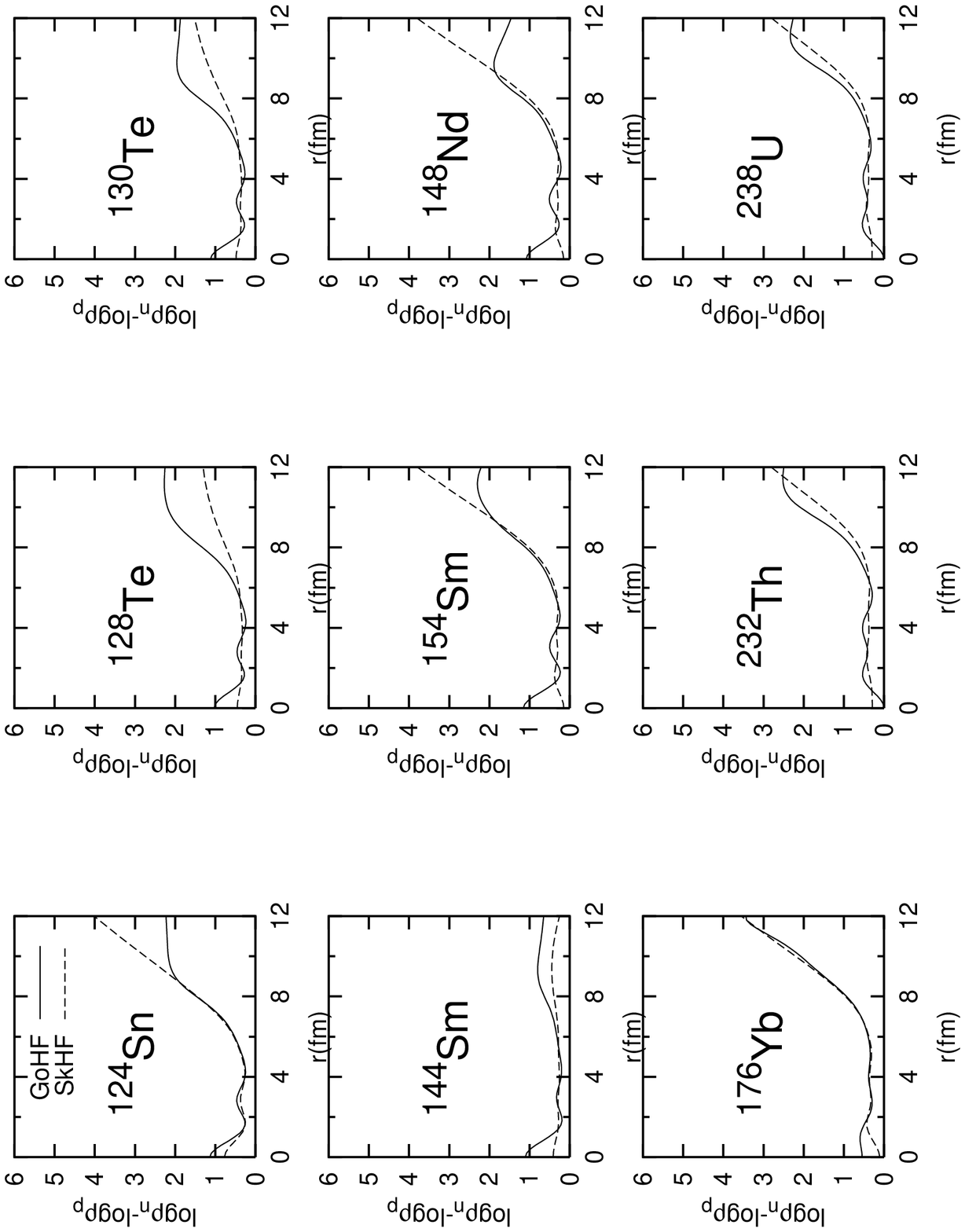,width=140mm}

\newpage

\framebox{Fig.~10a}

\vspace{0.5cm}
\epsfig{file=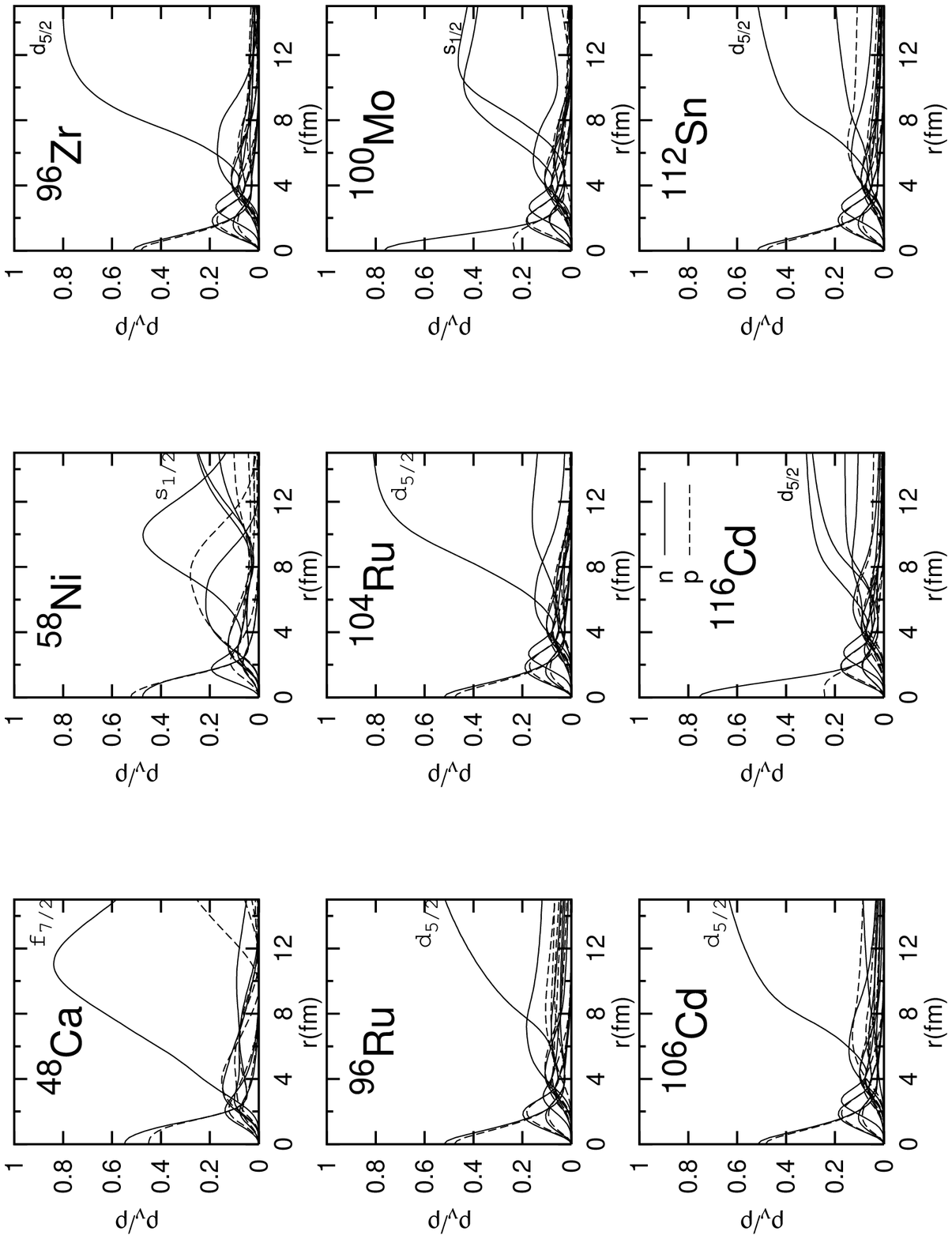,width=140mm}
\newpage

\framebox{Fig.~10b}

\vspace{0.5cm}
\epsfig{file=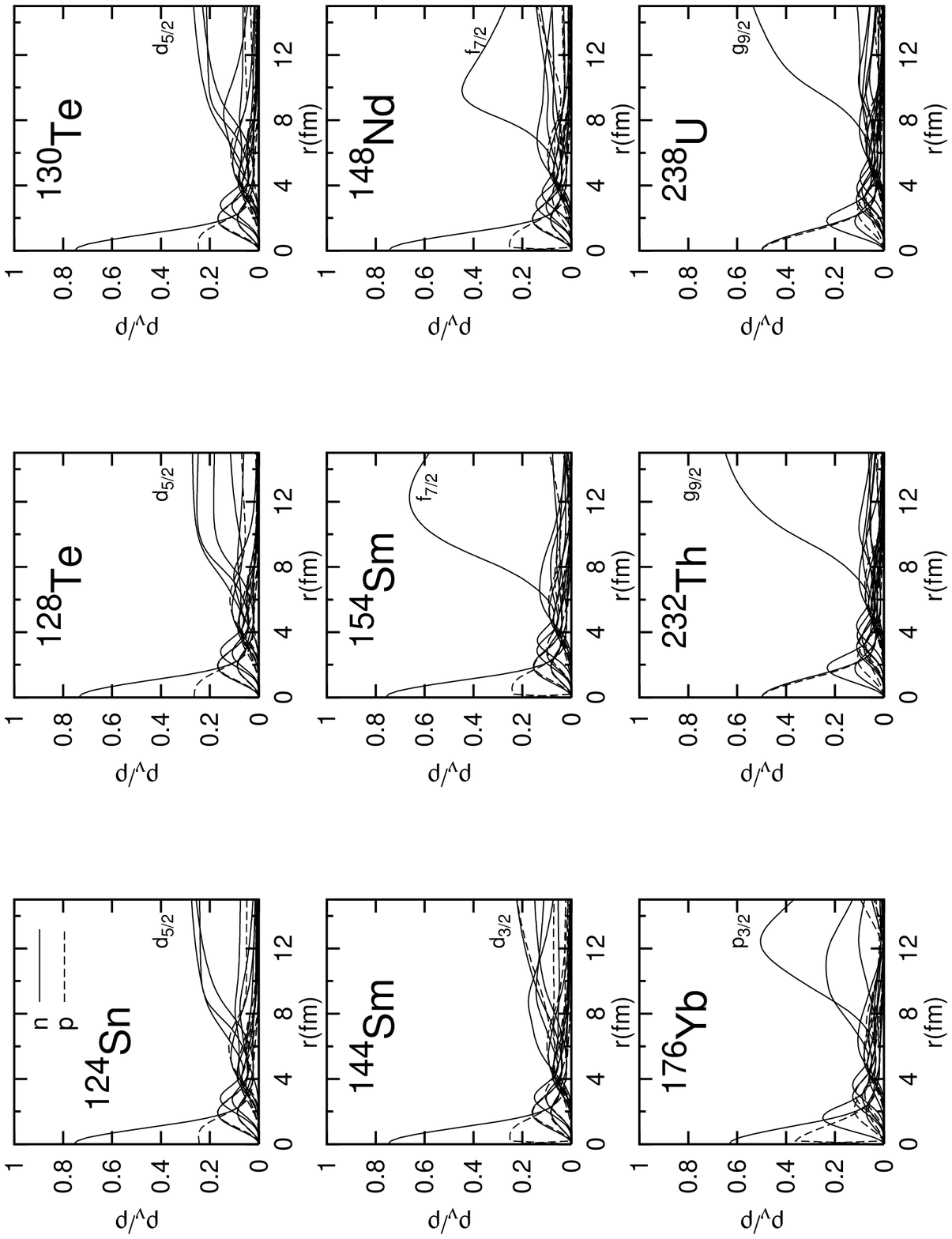,width=140mm}

\newpage

\framebox{Fig.~11a}

\vspace{0.5cm}
\epsfig{file=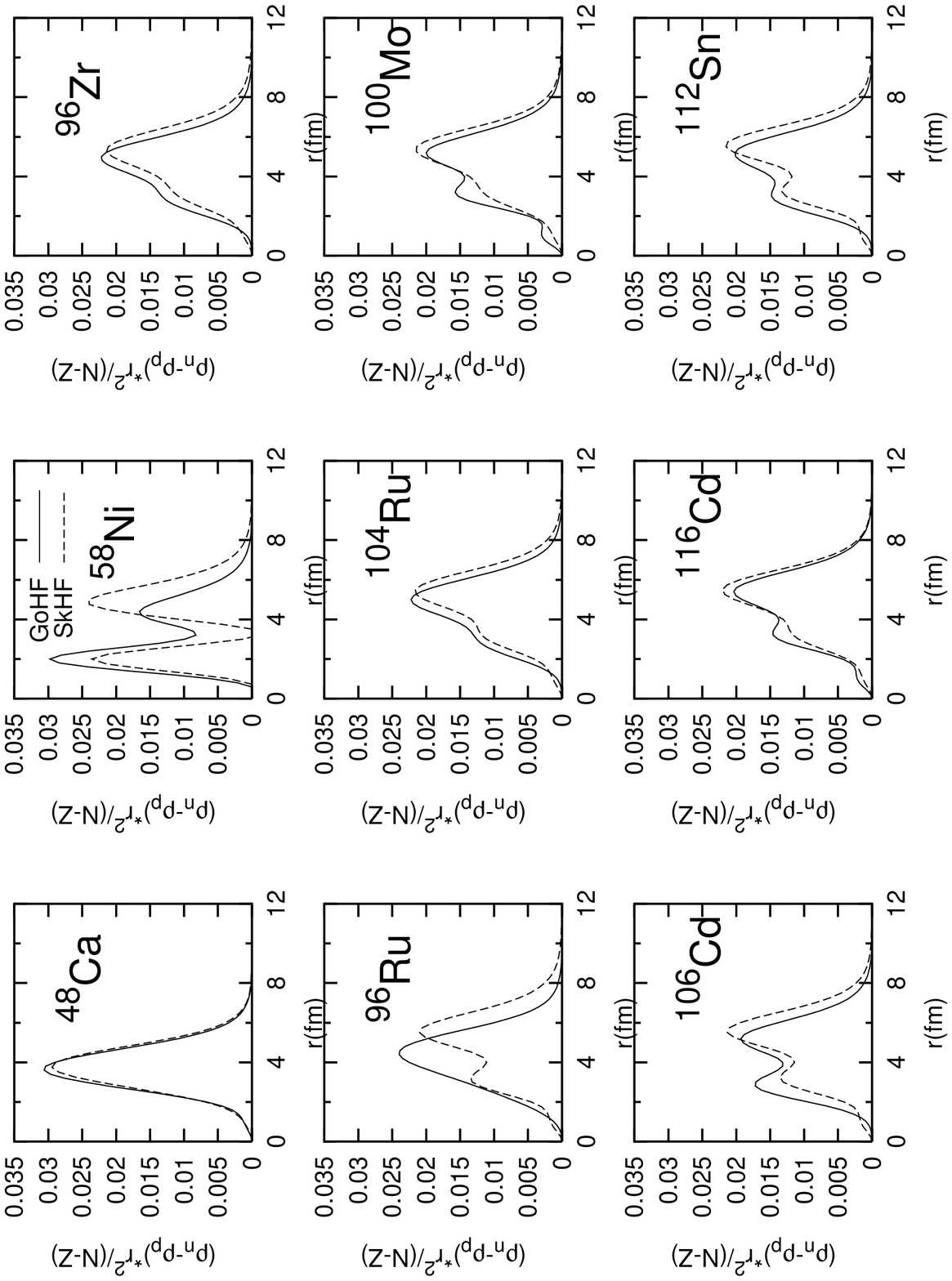,width=140mm}

\newpage

\framebox{Fig.~11b}

\vspace{0.5cm}
\epsfig{file=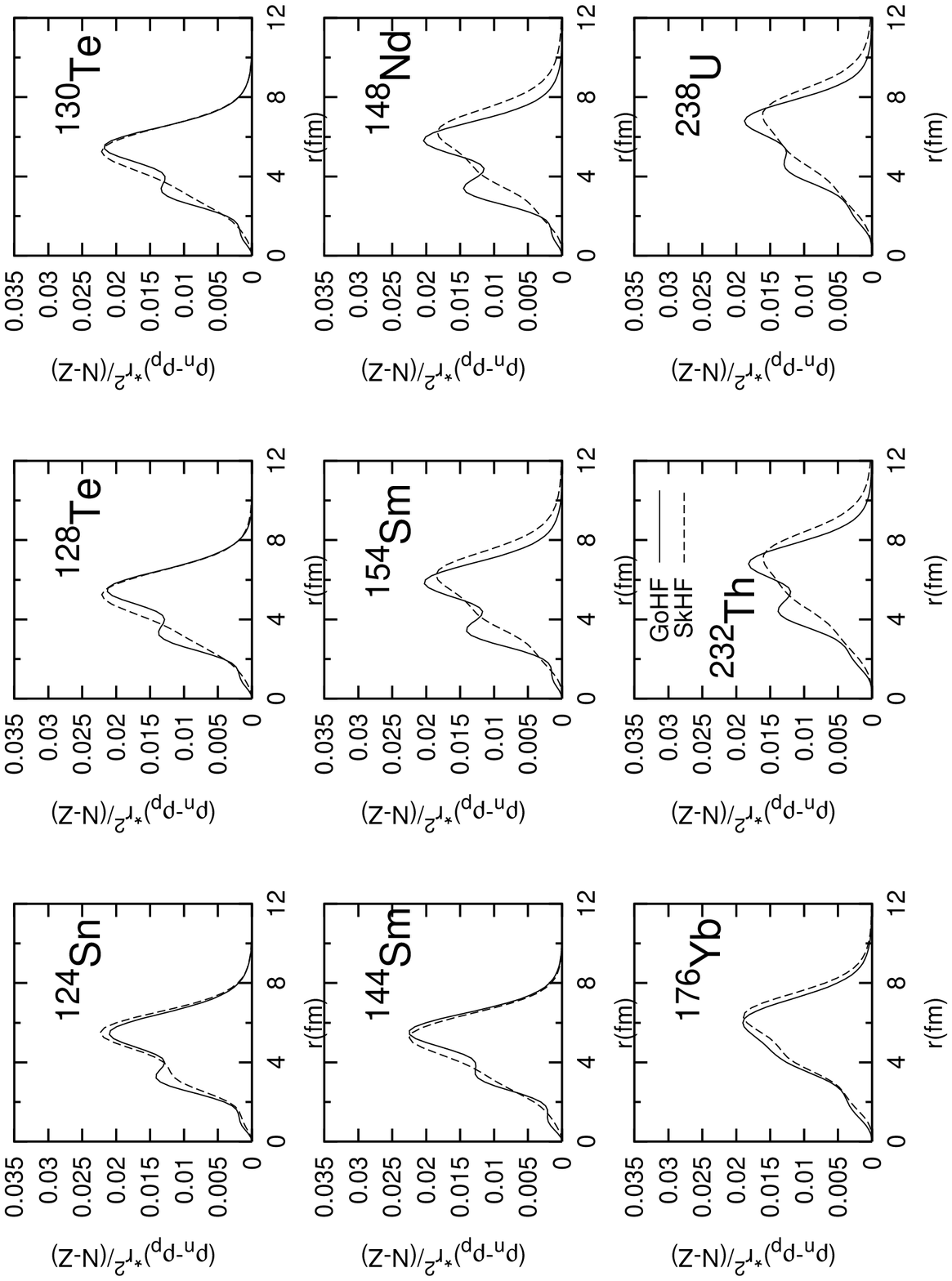,width=140mm}
\newpage

\framebox{Fig.~12a}

\vspace{0.5cm}
\epsfig{file=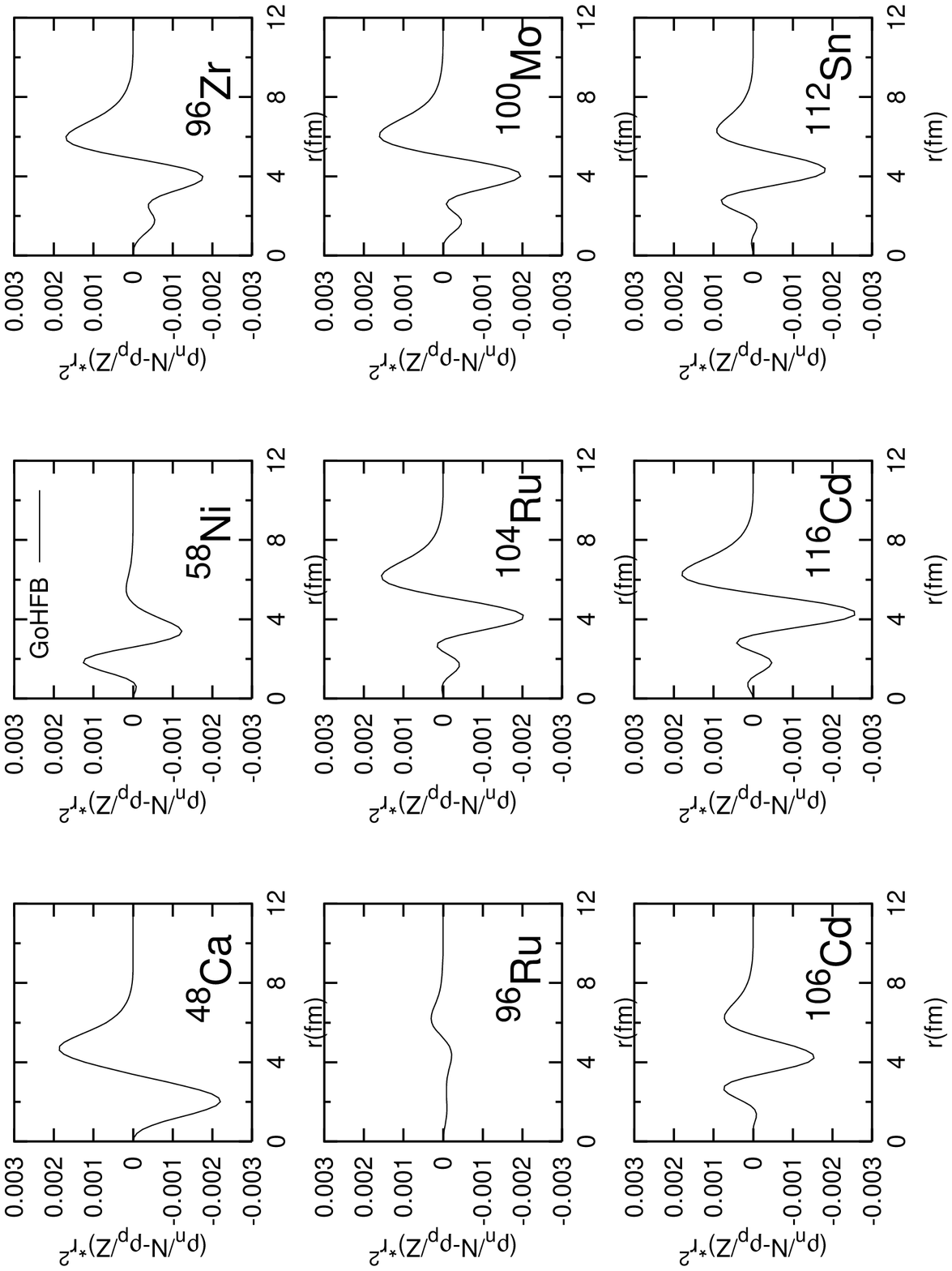,width=140mm}

\newpage

\framebox{Fig.~12b}

\vspace{0.5cm}
\epsfig{file=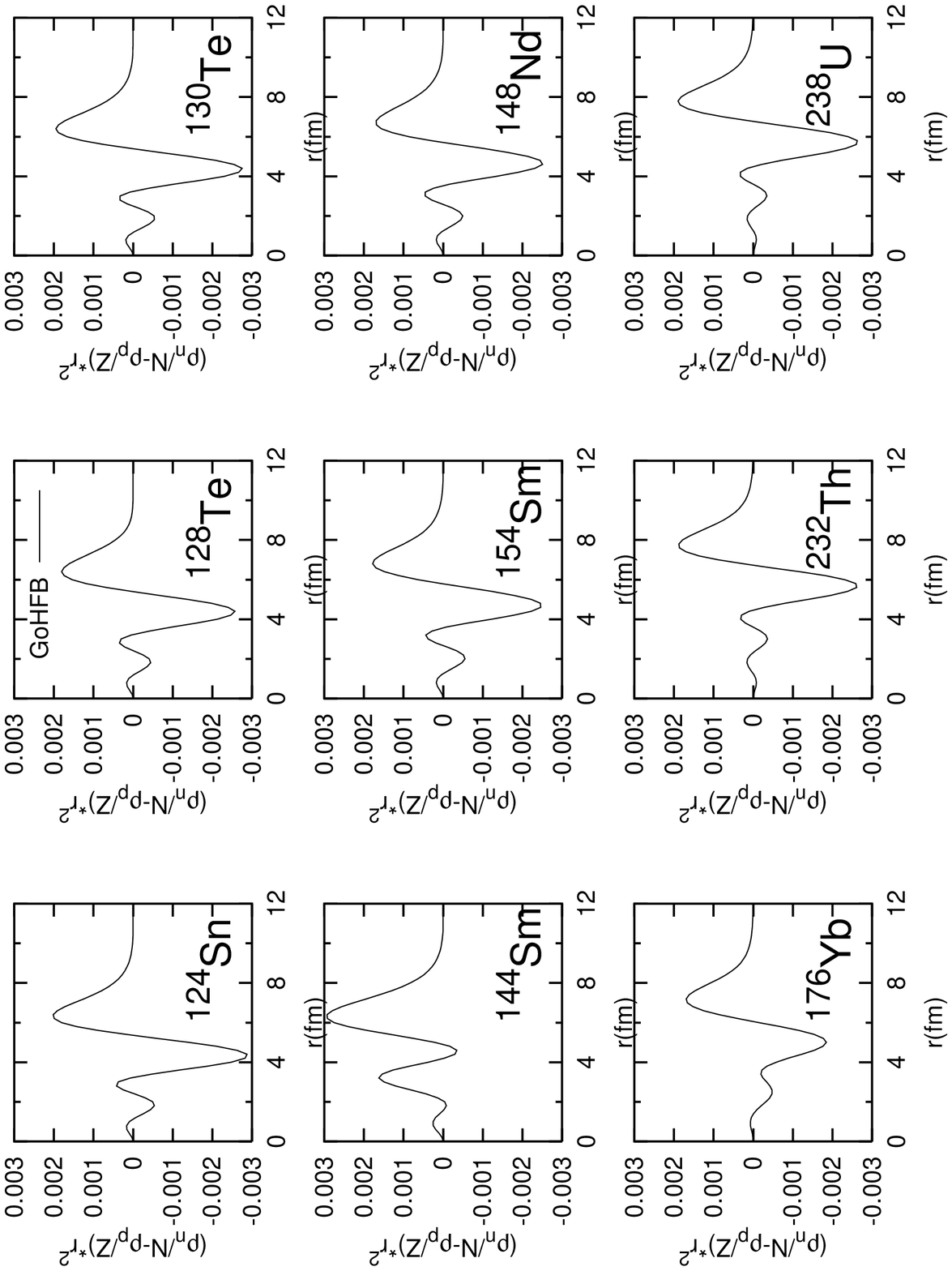,width=140mm}

\newpage

\framebox{Fig.~13}

\vspace{0.5cm}
\epsfig{file=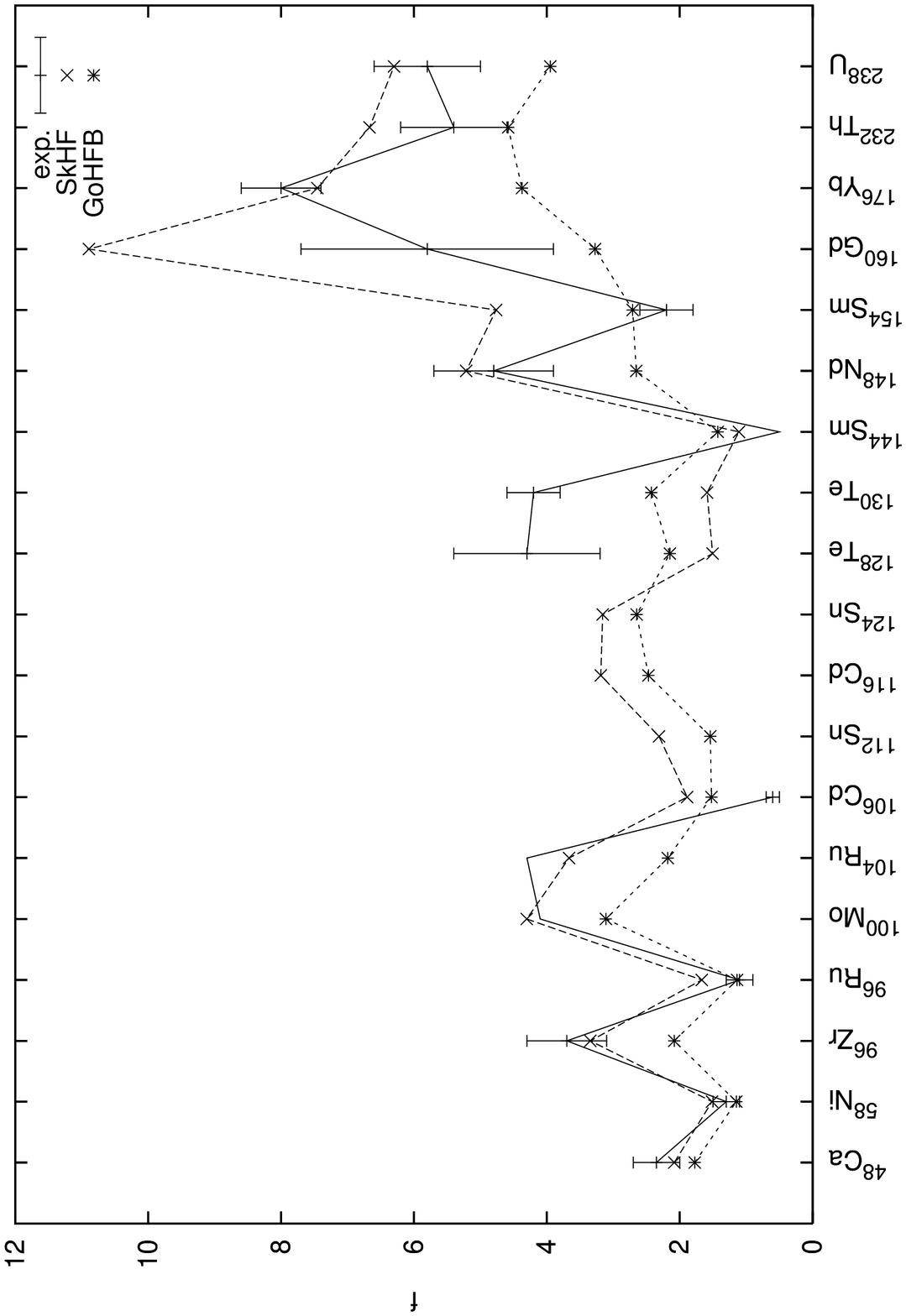,width=140mm}

\end{document}